\documentstyle[aps,psfig,multicol]{revtex}
\begin{document} 
\draft
\title{On evaluation of transverse spin fluctuations 
in quantum magnets}
\author{Avinash Singh}
\address{Department of Physics, Indian Institute of Technology, Kanpur-208016, India}
\maketitle
\begin{abstract} 
A numerical method is described for evaluating transverse 
spin correlations in the random phase approximation.
Quantum spin-fluctuation corrections to 
sublattice magnetization are evaluated for the antiferromagnetic
ground state of the half-filled Hubbard model in two
and three dimensions in the whole $U/t$ range.
Extension to the case of defects in the AF is also discussed
for spin vacancies and low$-U$ impurities. 
In the $U/t\rightarrow\infty$ limit, 
the vacancy-induced enhancement in the spin fluctuation correction 
is obtained for the spin-vacancy problem in two dimensions, for
vacancy concentration up to the percolation threshold. 
For low-$U$ impurities, the overall spin fluctuation correction 
is found to be strongly suppressed, although surprisingly 
spin fluctuations are locally enhanced at the low$-U$ sites. 

\end{abstract} 

\begin{multicols}{2}\narrowtext
\section{Introduction}
Transverse spin fluctuations are gapless, low-energy excitations in 
the broken-symmetry state of magnetic systems possessing
continuous spin-rotational symmetry. 
Therefore at low temperatures they play an important
role in diverse macroscopic properties
such as existence of long-range order, 
magnitude and temperature-dependence of the order parameter,
N\'{e}el temperature, spin correlations etc.
Specifically in the antiferromagnetic (AF) ground state
of the half-filled Hubbard model
transverse spin fluctuations are important both in 
two and three dimensions, where antiferromagnetic long-range 
order (AFLRO) exists at $T=0$. 
In the strong-coupling limit $(U/t\rightarrow \infty)$, 
where spin fluctuations are strongest,
they significantly reduce the zero-temperature 
AF order parameter in two dimensions to nearly $60\%$ of the
classical (HF) value.\cite{young,carlson,spfluc,quantum} 
Similarly the N\'{e}el temperature 
in three dimensions is reduced to nearly $65\%$ 
of the mean-field result $T_{\rm N}^{\rm MF}=zJ/4=6t^2/U$, for the equivalent
$S=1/2$ quantum Heisenberg antiferromagnet (QHAF).\cite{neel}

Recently there has also been interest in spin fluctuations 
due to defects, disorder and vacancies in the quantum antiferromagnet. 
In the random-$U$ model, where $U$ is set to zero
on a fraction $f$  of sites, the lattice-averaged AF order parameter
appears to be enhanced for small $f$, as seen in QMC calculations,\cite{denteneer}
presumably due to an overall suppression of quantum spin fluctuations.
On the other hand spin fluctuations are enhanced by strong disorder
in the Hubbard model with random on-site energies.
In the strong disorder regime, overlap of the two Hubbard bands 
leads to formation of essentially empty and doubly-occupied sites, 
which act like spin vacancies.\cite{dis} 
The problem of spin vacancies in the quantum antiferromagnet
is also relevant to the
electron-doped materials like ${\rm Nd_{2-x}Ce_x Cu O_4}$,
where spin-dilution behavior is observed.\cite{keimer1,manousakis}
While the problem of magnon 
energy renormalization due to spin vacancies has been addressed 
recently,\cite{bulut,brenig+kampf,dkumar,poilblanc,gapstate,sws}
these methods are limited to the low-concentration limit, and the 
vacancy-induced enhancement in transverse spin fluctuations 
has not been studied in the whole range of vacancy concentration.

In this paper we describe 
a new method for evaluating transverse spin correlations
and quantum spin-fluctuation corrections 
about the HF-level broken-symmetry state, in terms of
magnon mode energies and spectral functions obtained in the 
random phase approximation (RPA). 
The method is applicable in the whole $U/t$ range of interaction
strength, and is illustrated with three applications
involving the AF ground state of the half-filled Hubbard model --- 
(i) the pure model in $d=2,3$,
(ii) spin vacancies in the strong coupling limit in $d=2$, and 
(iii) low-$U$ impurities in $d=2$.
This method for obtaining quantum correction to sublattice magnetization
solely in terms of transverse spin correlations is parallel
to the spin-wave-theory (SWT) approach,\cite{anderson,oguchi}
and differs from the method involving
self-energy corrections.\cite{schrieffer}

The RPA approach has been demonstrated earlier to 
properly interpolate between the weak and strong coupling limits
for the spin-wave velocity.\cite{spfluc,schrieffer}
By going beyond the RPA level 
within a systematic inverse-degeneracy expansion scheme, 
which preserves the spin-rotational symmetry and hence the Goldstone
mode order by order, it was also shown that in the strong coupling limit
identical results are obtained for all quantum corrections, order
by order, as from the SWT approach for the QHAF.\cite{quantum} 
A renormalized RPA approach has also been devised recently
to obtain the magnetic phase diagram for the three dimensional
Hubbard model in the whole $U/t$ range,
and the $T_{\rm N}$ vs. $U$ behaviour was shown to properly
interpolate between the weak and strong coupling limits.\cite{neel}

\section{Transverse spin correlations}
The method is based on a convenient way to perform the frequency integral
in order to obtain spin correlations from spin propagators,
and we illustrate it here for transverse spin correlations.
We write the time-ordered, transverse spin propagator 
for sites $i$ and $j$, 
$\langle \Psi_{\rm G} | T [ S_i ^- (t) S_j ^+ (t')]|\Psi_{\rm G}\rangle$
at the RPA level in frequency space as,
\begin{equation}
[\chi^{-+}(\omega)]=\frac{[\chi^0(\omega)]}{1-U[\chi^0(\omega)]}
=\sum_n \frac{\lambda_n(\omega)}{1-U\lambda_n(\omega)} 
|\phi_n(\omega)\rangle \langle \phi_n(\omega) | \; ,
\end{equation}
where $|\phi_n(\omega)\rangle$ 
and $\lambda_n(\omega)$ are the eigenvectors
and eigenvalues of the $[\chi^0(\omega)]$ matrix. 
Here $[\chi^0(\omega)]_{ij}=i\int (d\omega'/2\pi)
G_{ij}^{\uparrow}(\omega')G_{ji}^{\downarrow}(\omega'-\omega)$
is the zeroth-order, antiparallel-spin particle-hole propagator,
evaluated in the self-consistent, broken-symmetry
state from the HF Green's functions $G^{\sigma}(\omega)$.
Spin correlations are then obtained from,
\begin{eqnarray}
\langle S^- _i (t) S^+ _j (t')\rangle_{\rm RPA}
&=&-i \int \frac{d\omega}{2\pi}
[\chi^{-+}(\omega)]_{ij}\; e^{-i\omega (t-t')} \nonumber \\
&=&\pm \sum_n \frac{\phi_n ^i (\omega_n)\phi_n ^j (\omega_n)}
{U^2 (d\lambda_n/d\omega)_{\omega_n}} 
e^{-i\omega_n (t-t')} \; ,
\end{eqnarray}
where the collective mode energies $\omega_n$ are obtained from 
$1-U\lambda_n(\omega_n)=0$, and 
$\lambda_n(\omega)$ has been Taylor-expanded as 
$\lambda_n(\omega) \approx \lambda_n(\omega_n) + (\omega-\omega_n) 
(d\lambda_n/d\omega)_{\omega_n}$ near the mode energies
to obtain the residues. 
For convergence, the retarded (advanced) part of the 
time-ordered propagator $\chi^{-+}$, having pole below (above) the
real-$\omega$ axis, is to be taken for $t' < t$ ($t' > t$).
The frequency integral is conveniently replaced by 
an appropriate contour integral
in the lower or upper half-plane in the complex-$\omega$ space
for these two cases, respectively, which results in Eq. (2).  

\section{Hubbard model in $d=$ 2,3}
We first illustrate this method for the half-filled Hubbard model 
in two and three dimensions 
on square and simple-cubic lattices, respectively. 
In this case it is convenient to use the two-sublattice representation
due to translational symmetry, and we work with 
the $2\times 2$ matrix $[\chi^0(q\omega)]$ in momentum space, 
which is given in terms of eigensolutions
of the HF Hamiltonian matrix.\cite{spfluc} 
The $k$-summation is performed numerically 
using a momentum grid with $\Delta k=0.1$ and 0.05, in three
and two dimensions, respectively. 

Equal-time, same-site 
transverse spin correlations are then obtained from Eq. (2)
by summing over the different $q$ modes, 
using a momentum grid 
with $\Delta q=0.3$ and $0.1$ in three and two dimensions, respectively.
We consider $t'\rightarrow t^-$, so that the retarded part is used, with
positive mode energies. 
From spin-sublattice symmetry, correlations on A and B sublattice sites 
are related via 
$\langle S^+ S^- \rangle_{A}= \langle S^- S^+ \rangle_{B}$.
Thus the transverse 
spin correlations are obtained from magnon amplitudes on A and B sublattices,
and from Eq. (2) we have
\begin{eqnarray}
\langle S^- S^+ \rangle_{\rm RPA} &=& 
\sum_q \frac{(\phi_q ^A)^2}
{U^2 (d\lambda_q/d\omega)_{\omega_q}} \nonumber \\
\langle S^+ S^- \rangle_{\rm RPA} &=& 
\sum_q \frac{(\phi_q ^B)^2}
{U^2 (d\lambda_q/d\omega)_{\omega_q}}  \; .
\end{eqnarray}

From the commutation relation $[S^+,S^-]=2S^z$,  
the difference $\langle S^+ S^- - S^- S^+ \rangle_{\rm RPA}$,
of transverse spin correlations evaluated at the RPA level, 
should be identically equal to $\langle 2S^z \rangle_{\rm HF}$.
This is becasue both the RPA and HF approximations are
O(1) within the inverse-degeneracy expansion scheme\cite{quantum}
in powers of $1/{\cal N}$ (${\cal N}$ is the number of orbitals per site),
and therefore become exact in the limit ${\cal N} \rightarrow\infty$,
when all corrections of order $1/{\cal N}$ or higher vanish.
This is indeed confirmed as shown in Figs. 1 and 2. 
The deviation at small $U$ is because 
of the neglect in Eq. (2) of the 
contribution from the single particle excitations
across the charge gap,
arising from the imaginary part of $\chi^0(\omega)$ in Eq. (1).

The sum $\langle S^+ S^- + S^- S^+ \rangle_{\rm RPA}$
yields a measure of transverse spin fluctuations about the HF state,
and in the strong coupling for spin $S$, one obtains 
$\langle S^+ S^- + S^- S^+ \rangle_{\rm RPA}
=(2S)\sum_q 1/\sqrt{1-\gamma_q^2}$.\cite{quantum,anderson}
Using the identity,
$\langle S^z S^z\rangle =S(S+1) - \langle S^+ S^- + S^- S^+ \rangle/2$
in this limit, 
the sublattice magnetization $m=\langle 2S^z\rangle$ 
is then obtained from,
\begin{equation}
\langle S^z\rangle =
S\left [1- \frac{1}{S} \left ( \frac{\langle S^+ S^- + S^- S^+ \rangle}
{2S} -1 \right ) \right ]^{1/2} .
\end{equation}
To order $1/2S$, this yields the 
correction to the sublattice magnetization 
of $(2S)^{-1}\sum_q [(1-\gamma_q^2)^{-1/2} -1]=0.156$ and 0.393
for $S=1/2$ in three  and two dimensions, 
respectively. 
This same result at one loop level was also obtained 
from a different approach in terms of the electronic spectral weight
transfer,\cite{spfluc,quantum} 
and is in exact agreement with the SWT result.\cite{anderson,oguchi}

As $\langle S^z\rangle_{\rm HF}$ is the maximum (classical) 
spin polarization in the z-direction,
and therefore also the maximum eigenvalue of the 
local $S_z$ operator,
therefore for arbitrary $U$, 
the HF magnitude $\langle S^z\rangle_{\rm HF}$ 
plays the role of the effective spin quantum number, $S$.
The sublattice magnetization $m$ is therefore obtained 
from $m=m_{\rm HF} -\delta m_{\rm SF}$, 
where the first-order, quantum spin-fluctuation correction 
$\delta m_{\rm SF}$ is obtained from Eq. (4) with 
$S=\langle S^z\rangle_{\rm HF}$,
\begin{equation}
\delta m_{\rm SF}=
\frac{\langle S^+ S^- +  S^- S^+ \rangle_{\rm RPA} }
{\langle S^+ S^- -  S^- S^+ \rangle_{\rm RPA} } -1 .
\end{equation}
In the strong coupling limit $U/t\rightarrow\infty$, 
$\langle S^+ S^- -  S^- S^+ \rangle_{\rm RPA} =
\langle 2S^z\rangle_{\rm HF} =1$
for $S=1/2$, so that the spin-fluctuation correction simplifies to, 
$\delta m_{\rm SF}=2\langle S^- S^+ \rangle_{\rm RPA}$.
For a site on the B sublattice, where 
$\langle S^+ S^- -  S^- S^+ \rangle_{\rm RPA} =
\langle 2S^z\rangle_{\rm HF} =-1$, we have 
$\delta m_{\rm SF}=2\langle S^+ S^- \rangle_{\rm RPA}$.

\begin{figure}
\vspace*{-60mm}
\hspace*{-28mm}
\psfig{file=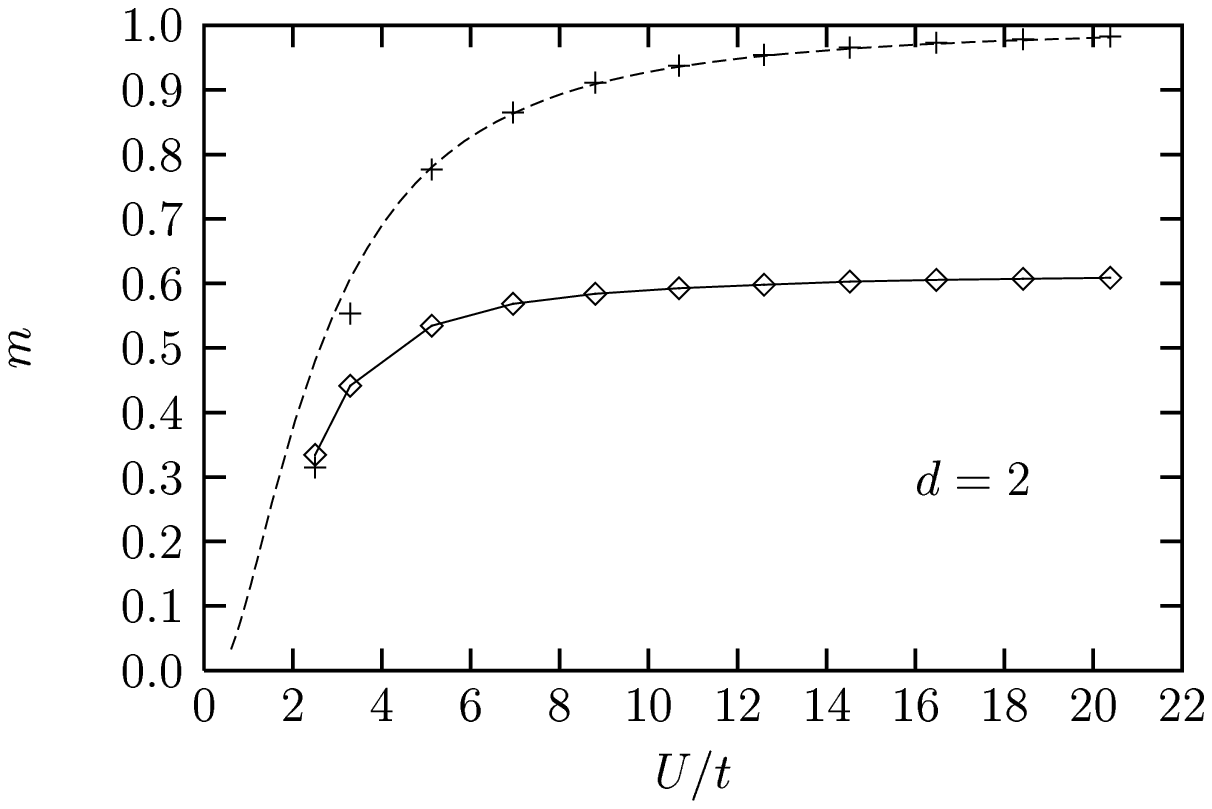,width=135mm,angle=0}
\vspace{-70mm}
\caption{The sublattice magnetization $m$ vs. $U$ in two
dimensions (diamonds), along with 
the HF results from (i) the self-consistency condition (dashed),
and (ii) $\langle 2S_z\rangle_{\rm HF}=
\langle S^+S^- - S^-S^+\rangle_{\rm RPA}$ (plus).}
\end{figure}

\begin{figure}
\vspace*{-60mm}
\hspace*{-28mm}
\psfig{file=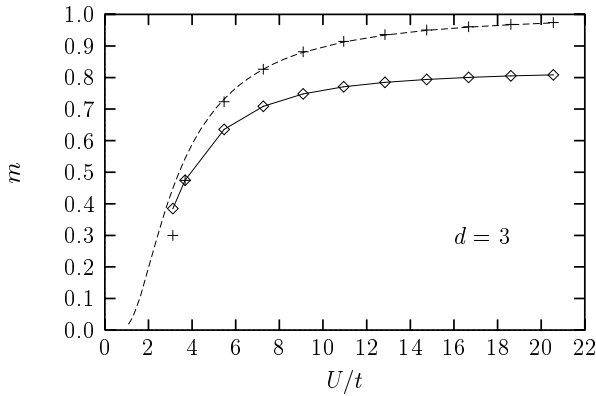,width=135mm,angle=0}
\vspace{-70mm}
\caption{The sublattice magnetization $m$ vs. $U$ in three
dimensions (diamonds), along with 
the HF results from (i) the self-consistency condition (dashed),
and (ii) $\langle 2S_z\rangle_{\rm HF}=
\langle S^+S^- - S^-S^+\rangle_{\rm RPA}$ (plus).}
\end{figure}

The $U$-dependence of sublattice magnetization $m$ 
is shown in Figs. 1, and 2 for $d=2$ and $d=3$, respectively.
In both cases
it interpolates properly between the weak and strong coupling limits,
approaching the SWT results 0.607 and 0.844,
respectively, as $U/t\rightarrow\infty$.
A comparison of the $m$ vs. $U$ behaviour with earlier results 
is presented in Fig. 3 for the well studied $d=2$ case.
Earlier studies have employed a variety of methods including 
the variational Monte Carlo (VMC),\cite{vmc}
self-energy corrections (SE),\cite{schrieffer}
quantum Monte Carlo (QMC),\cite{hirsch.tang}
functional-integral schemes,\cite{hasegawa}
the generalized linear spin-wave approximation (GLSWA),\cite{mehlig}
and mapping of low-energy excitations to those of a QHAF 
with $U$-dependent, extended-range spin couplings.\cite{logan.book}

\begin{figure}
\vspace*{-60mm}
\hspace*{-28mm}
\psfig{file=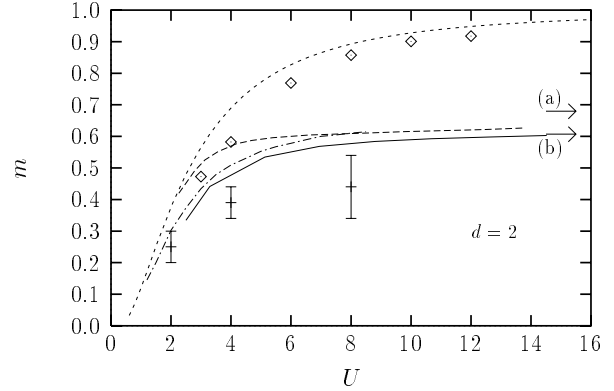,width=135mm,angle=0}
\vspace{-70mm}
\caption{The sublattice magnetization $m$ vs. $U$ 
from our RPA spin-fluctuation approach (solid line)
compared with earlier results --- 
VMC [19] (diamonds),
SE [18] (dash-dot), 
QMC [20] (errorbars), and 
GLSWA [22] (dash). 
Also shown is the HF result (dotted).
The arrows denote the asymptotic results of
(a) [2] and (b) [1].}
\end{figure}

In addition to the two-sublattice basis,
we have also used the full site representation 
in the strong coupling limit, in order to illustrate the
scaling of the quantum correction with system size.
Here the $\chi^0$ matrix is evaluated and diagonalized in the 
site basis for finite lattices.
Results for lattice sizes with 
$8\le L\le 16$ are shown in Fig. 4.
A quadratic least-square fit is
used to extrapolate to infinite system size, which yields 
$\delta m_{\rm SF}(1/L\rightarrow 0)=0.39$, in agreement
with the result from Eq. (4). 

\section{Spin vacancies}
The site representation has also been used to
obtain transverse spin fluctuations for the problem of spin
vacancies in the AF in the limit $U/t\rightarrow\infty$. 
As mentioned already this method is applicable 
in the whole range of vacancy concentration, and allows determination 
of the critical vacancy concentration at which the AF order parameter vanishes. 
For the vacancy problem we consider the following Hamiltonian on a square lattice
with nearest-neighbor (NN) hopping,
\begin{equation}
H=-\sum_{<ij>\sigma} t_{ij} (a_{i\sigma}^{\dagger} a_{j\sigma} +{\rm h.c.})
+ U\sum_i \hat{n}_{i\uparrow}\hat{n}_{i\downarrow} \; ,
\end{equation}
where the hopping terms $t_{ij}=0$ if sites $i$ or $j$ are vacancy sites,
and $t_{ij}=t$ otherwise. Thus, for a vacancy on site $i$,
all hopping terms $t_{ij}$ connecting $i$ to its NN sites $j$ 
are set to zero.
The vacancy site is thus completely decoupled from the system.
Half-filling is retained by having one fermion per remaining site.
We consider the $U/t\rightarrow\infty$ limit,
where the local moments are fully saturated,
and  the vacancy problem becomes identical to the 
\begin{figure}
\vspace*{-60mm}
\hspace*{-28mm}
\psfig{file=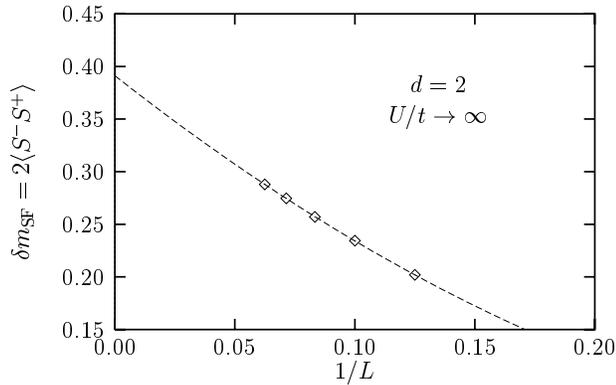,width=135mm,angle=0}
\vspace{-70mm}
\caption{The spin-fluctuation correction to sublattice magnetization 
$\delta m_{\rm SF}$ vs. $1/L$, for $L=$ 8, 10, 12, 14, 16.}
\end{figure}
\noindent
spin-vacancy problem in the QHAF. 
This is also equivalent to the problem of non-magnetic impurities in
the AF in the limit of the 
impurity potential $V\rightarrow\infty$.\cite{gapstate,sws}

The structure of the $\chi^0(\omega)$ matrix in the host AF, and 
the modification introduced by spin vacancies has been considered 
earlier in the  context of static impurities.\cite{sws}
As the Goldstone mode is preserved, it is convenient to work with the  
matrix $K\equiv U[1-U\chi^0]$, inverse of which 
yields the spin propagator in Eq. (1) near the mode energies.
When expressed in units of
$U^2t^2/\Delta^3 =2J$, where $2\Delta=m_{\rm HF} U$ is the AF gap
parameter, and $J=4t^2/U$,
it has the following simple structure for the host AF:
\begin{eqnarray}
K^0_{ii} &=& 1+\omega \;\;\;\;\;i\;{\rm in\; A \; sublattice} \nonumber \\
K^0_{ii} &=& 1-\omega \;\;\;\;\;i\;{\rm in\; B \; sublattice} \nonumber \\
K^0_{ij} &=& 1/4  \;\;\;\;\;\;\; j\; {\rm is\; NN\; of \;} i. 
\end{eqnarray}

Vacancies introduce a perturbation in $K$ due to absence of
hopping between the vacancy and NN sites, and we take 
$\delta K \equiv -U^2\delta \chi^0$
to refer to this vacancy-induced perturbation, 
so that $K=K^0 +\delta K$. If site $i$ is a vacancy site 
and $j$ the NN sites, then 
the matrix elements of $\delta K$ are,
\begin{equation}
\delta K_{ij}=\delta K_{ji}=\delta K_{jj} =  -1/4 \; ,
\end{equation}
and the magnitude of $K_{ii}$ is irrelevant since the vacancy site
$i$ is decoupled. 
Thus $K_{ij}=0$, reflecting the decoupling of the vacancy, 
and the static part of the diagonal
matrix elements $K_{jj}$ on NN sites are reduced by 1/4. 
For $n$ vacancies on NN sites, the static part is 
$1-n/4$. This ensures that the Goldstone mode is preserved.
Thus if a spin on site $j$ were surrounded by a maximum 
of four vacancies on NN sites,
then the static part vanishes, and $K_{jj}=\pm \omega$,
representing an isolated spin, which yields a $1/\omega$ pole
in the transverse spin propagator. This is an isolated single-spin cluster,
and with increasing vacancy concentration, larger isolated spin clusters are formed.
\begin{figure}
\vspace*{-60mm}
\hspace*{-28mm}
\psfig{file=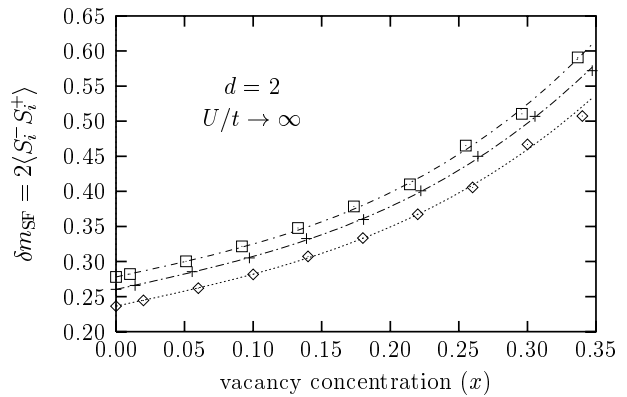,width=135mm,angle=0}
\vspace{-70mm}
\caption{The spin-fluctuation correction to sublattice magnetization
$\delta m_{\rm SF} $ vs. $x$, 
for $L=10$ (diamonds), 12 (plus), and 14 (squares).}
\end{figure}
\noindent
As these are decoupled from the remaining system, their spin-fluctuation
contributions are not included, as the OP vanishes for finite spin clusters.
When $x$ exceeds the percolation threshold $\sim 0.4$, the fraction
of macroscopically large spin clusters in the system vanishes, and therefore
no AFLRO is possible for $x>0.4$. 

For a given vacancy concentration and system size, the appropriate number
of vacancies are placed randomly across the lattice, and the matrix $K$ 
constructed accordingly. 
Exact diagonalization of $K$ is carried out, and
the eigensolutions are used to compute the transverse spin correlations from Eq. (2).
The quantum, spin-fluctuation correction is then obtained from 
the strong-coupling limit of Eq. (5).
The transverse spin correlation 
$\langle S^- S^+ \rangle$ is averaged over all spins within the 
A sublattice. As mentioned already, only the spins in the macroscopic
cluster spanning the whole lattice are considered, and contributions from
spins in isolated spin clusters are excluded. Configuration
averaging over several realizations of the vacancy distribution is also
carried out. 

The quantum spin-fluctuation correction vs. vacancy concentration 
is shown in Fig. 5 for three lattice sizes, $L=10,12,14$.
Best fits are obtained with an expression including a cubic term,
$\delta m_{\rm SF}=\alpha+\beta x +\gamma x^3$. For the three
lattice sizes $\alpha=$0.236, 0.260 and 0.278, respectively, 
and as shown in Fig. 3,
it extrapolates to 0.39 as $1/L\rightarrow 0$. 
The coefficient of the linear term is
found to be nearly independent of system size, $\beta \approx 0.42$.
And the cubic term $\gamma$ takes values approximately 
3.6, 4.0, and 4.4 for the three lattice sizes, 
and extrapolates to about 6.5 as $1/L\rightarrow 0$. 
With these coefficients, the spin-fluctuation correction 
$\delta m_{\rm SF}$ is nearly 1 for $x=0.4$.
Beyond $x=0.4$, the percolation limit,
there is no single, macroscopically large 
spin cluster left in the system.
Therefore the point where the AF order parameter vanishes
and AFLRO is destroyed due to spin fluctuations nearly
coincides with the percolation threshold. 
This is in agreement with results from series expansion\cite{series}
and quantum Monte Carlo simulations\cite{behre} 
of the QHAF with spin vacancies.

\section{Random-$U$ model}
We now consider a quenched impurity model with a random
distribution of impurity sites characterized by
a local Coulomb interaction $U'\ne U $ for the host sites. 
With H and I
referring to the sets of host and impurity sites
respectively, we consider the following Hubbard model
in the particle-hole symmetric form
at half-filling and on a square lattice,
\begin{eqnarray}
H &=& -t \sum_{<ij>\sigma} (a_{i\sigma}^{\dagger}a_{j\sigma}
+ {\rm h.c.} ) +
U \sum_{\rm i \in H} 
(n_{i\uparrow}-\frac{1}{2})
(n_{i\downarrow}-\frac{1}{2})\nonumber \\
&+&
U' \sum_{\rm i \in I} 
(n_{i\uparrow}-\frac{1}{2})
(n_{i\downarrow}-\frac{1}{2})  .
\end{eqnarray}

The motivations for studying this impurity model are threefold. 
In view of the observed {\em enhancement} of magnetic 
order at low concentration of impurities,\cite{denteneer}
we shall analyze the suppression of quantum spin fluctuations
to examine whether this is due to a local 
suppression at the low-$U$ sites. 
The RPA evaluation of transverse spin correlations is also 
extended to the case of site-dependent interactions.
Furthermore, at half-filling this model also provides a simplistic 
representation for magnetic impurity doping in an AF.
This may appear contradictory in view of the 
apparently nonmagnetic ($U' \approx 0$) 
nature of the impurity sites. 
However, this feature is expressed only away from 
half-filling.\cite{denteneer} 

The atomic limit $t=0$ provides a convenient starting point
for further discussions. 
In the particle-hole symmetric form of Eq. (9), 
since not only local interaction terms, 
but the on-site energy terms are also modified 
(from $-U/2$ to $-U'/2$) at the impurity sites,
therefore the energy levels for
added hole and particle are $(-U/2,U/2)$ and $(-U'/2,U'/2)$ 
for host and impurity sites respectively.
To order $t^2$, this impurity model therefore canonically
maps to the following $S=1/2$ Heisenberg model
\begin{equation}
H_{\rm eff} = J\sum_{<ij>\in H} (\vec{S}_i . \vec{S}_j -\frac{1}{4})
+ J'\sum_{i\in I} \sum_\delta 
(\vec{S}_i . \vec{S}_{i+\delta} -\frac{1}{4})
\end{equation}
where in the first term $J=4t^2/U$ is the conventional exchange
coupling between neighboring host spins, and $J'=8t^2/
(U+U')$ is the exchange coupling between impurity spins and
neighboring host spins. In writing Eq. (10) we have assumed the
dilute impurity limit, and discounted the possibility of two 
impurity spins occupying NN positions, in which case the
impurity-impurity exchange coupling will be $4t^2/U'$.

Therefore, in the strong correlation limit, 
this random-$U$ model also describes
magnetic impurities in the AF within an impurity-spin model.
The magnetic-impurity doping is characterized by identical 
impurity and host spins $(S'=S)$, 
but with different impurity-host exchange coupling
$(J' \ne J)$. Both cases $J'>J$ or $J' < J$ are possible, and 
in this paper we have considered the two cases: (i) $U' << U$ so that
$J' \approx 2J$, and $U'=3U$ so that $J' = J/2$.
While this model is easily generalized to 
other magnetic-impurity models represented by 
locally modified impurity-host hopping terms $t' \ne t$, and/or 
different impurity energy levels,
in fact, the essential features are
already contained here as the impurity exchange coupling
$J'$ is the relevant quantity in
determining the spin fluctuation behavior.

\subsection*{RPA  with site-dependent interaction}
We recast the RPA expression for the transverse spin propagator
in a form suitable for site-dependent interactions.
In terms of a diagonal interaction matrix $[U]$, with elements 
$[U]_{ii}=U_i $, the local Coulomb interaction at site $i$,
the time-ordered transverse spin propagator at the RPA level 
can be rewritten, after simple matrix manipulations, as
\begin{equation}
[\chi^{-+}(\omega)]=
\frac{[\chi^{0}(\omega)]}
{1 - [U] [\chi^{0}(\omega)]} =
\frac{1}{[A(\omega)]} - \frac{1}{[U]}
\end{equation}
where $[A(\omega)]=[U] - [U][\chi^0 (\omega)][U]$ is a
symmetric matrix.
As $[U]$ is non-singular, the singularities in 
$[\chi^{-+}(\omega)]$, which yield the magnon modes,
are then given completely by the vanishing of the 
eigenvalues of the matrix $[A]$. In terms of
$\lambda_n$ and $|\phi _n \rangle $, the eigenvalues and
eigenvectors of the matrix $[A]$, we have  
$[A(\omega)]^{-1} = \sum_n \lambda_n(\omega)^{-1}
|\phi _n (\omega) \rangle \langle |\phi _n (\omega) |$,
so that the magnon-mode energies $\omega_n$ are then given by
$\lambda_n (\omega_n) =0$, 
and in analogy with Eq. (2) the transverse spin correlations 
are obtained from,
\begin{equation}
\langle S_i ^- (t) S_j ^+(t') \rangle_{\rm RPA} =
\pm \sum _n 
\frac{\phi_n ^i (\omega_n)
\phi_n ^j (\omega_n)}
{(d\lambda_n/d\omega)_{\omega_n}}
\;\;
e^{i\omega_n (t-t')}   .
\end{equation}

As we are interested in the dilute behaviour, we examine
the correction to sublattice magnetization due to
two impurities, one on each sublattice for symmetry.
Since $m=m_{\rm HF}-\delta m_{\rm SF}$, 
corrections to both $m_{\rm HF}$ and $\delta m_{\rm SF}$ are
expressed in the dilute limit (impurity concentration $x$) as  
$m_{\rm HF}=m_{\rm HF}^{0}-\alpha_{\rm HF} x$,
and $\delta m_{\rm SF}=\delta m_{\rm SF}^{0} -\alpha_{\rm SF} x$.
The overall $m(x)$ behavior 
therefore depends on the relative magnitude of the
coefficients $\alpha_{\rm HF}$ and $\alpha_{\rm SF}$.  
For $U=10$ and $U'=2$ we find, at the HF level, 
that $m_{\rm HF}^{0}=0.93$
and $\alpha_{\rm HF}=0.19$, both quite independent of system size.

The (site-averaged) spin fluctuation correction $\delta m_{\rm SF}$ 
is obtained from Eq. (5) with and without impurities,
and the impurity contribution extracted. 
For $U'=2$ we find a net {\em reduction} in $\delta m_{\rm SF}$.
Divided by $2/N$, the impurity concentration, this yields
the coefficient $\alpha_{\rm SF}$ defined above,
and also the per-impurity contribution to the total spin 
fluctuation correction over the whole lattice. 
The spin-fluctuation correction $\delta m_{\rm SF}$
for the pure case,
and the per-impurity contribution $(\alpha_{\rm SF})$
are shown in Fig. 6 for different lattice sizes,
along with least-square fits.
It is seen that in the infinite size limit,
the per-impurity reduction is nearly 0.2,
which is more than half of the correction per site 
in the pure case (0.35).
Thus, there is a substantial reduction in the
averaged spin fluctuation correction due to the low$-U$ impurities.

As the two coefficients 
$\alpha_{\rm HF}$ and $\alpha_{\rm SF}$ are very nearly the same,
the sublattice magnetization 
\begin{equation}
m(x)=m(0) - (\alpha_{\rm HF} -\alpha_{\rm SF})x
\end{equation}
shows negligible concentration dependence.
Thus the (relatively small) 
reduction in the HF value due to the low-$U$ impurities
is almost fully compensated by the (relatively substantial) 
reduction in the spin fluctuation correction.
To first order in $x$, we thus find that there is 
nearly no loss of AF order due to the low-$U$ impurities.
As mentioned earlier, even a slight enhancement
in the AF order was recently seen for the case $U'=0$.\cite{denteneer}

We next examine the site-dependence of the 
local spin fluctuation corrections $\delta m_{\rm SF}^i$
near the impurities.
Table I shows that spin fluctuation is
actually {\em enhanced} on the low-$U$ impurity sites.
The suppression of $\delta m_{\rm SF}^i$
in the vicinity more than compensates for this local enhancement,
resulting in an overall reduction on the average. 
On the other hand, for $U'=30$,
we find that the correction is suppressed on the high-$U$
impurity site, while it is enhanced on the average. 
Thus, to summarize, 
when the impurity spin is coupled more strongly (weakly), 
the spin-fluctuation correction is enhanced (suppressed) 
locally at the impurity site, 
but the average correction to 
sublattice magnetization is suppressed (enhanced).
\begin{figure}
\vspace*{-60mm}
\hspace*{-28mm}
\psfig{file=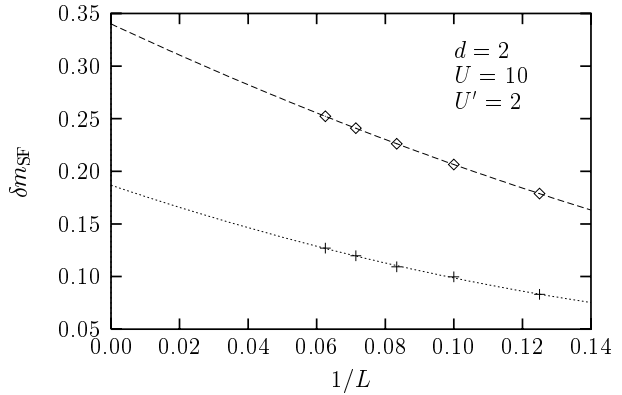,width=135mm,angle=0}
\vspace{-70mm}
\caption{The per-impurity reduction in the spin-fluctuation correction 
to sublattice magnetization $\delta m_{\rm SF} $ vs. $1/L$ (plus), 
for finite lattices with $L=8,10,12,14,16$.
Also shown is the correction per site for the pure system (diamonds),
indicating significant relative impurity contribution.}
\end{figure}

This local enhancement can be understood in terms of the 
correlations $\langle S_i ^- S_i ^+ \rangle$  as follows.
Since the impurity spin  
is more strongly coupled to the neighboring spins,
the NN  matrix elements $A_{i\delta}$ are enhanced.
This puts more magnon amplitude $\phi$ on the impurity site,
so that from Eq. (2) the transverse spin correlations
$\langle S_i ^- S_i ^+ \rangle$,
and therefore the spin-fluctuation correction,
are enhanced for low-$U$ impurities. 
The overall decrease in the averaged fluctuation correction,
however, is due to the stiffening of the magnon spectrum in the 
important low-energy  sector, 
following from the increased average spin coupling.
\end{multicols}
\widetext
\begin{table}
\caption{The local spin-fluctuation corrections $\delta m_{\rm sf} ^i $
for a $16 \times 16$ system ($U=10$), 
with two impurities at (11,4) and (4,14). 
For the two cases $U'=2$ ($J'\approx 2J$) and $U'=30$ ($J'=J/2$),
quantum corrections are enhanced/suppressed locally
at the impurity sites (indicated in boldfaces), 
but are suppressed/enhanced on the average.}
\begin{tabular}{ccccccccccccccccc}
site & 1 & 2 & 3 & 4 & 5 & 6 & 7 & 8 & 9 & 10 & 11 & 12 & 13 & 14 & 15 & 16 \\ \tableline
1 & & & & & & & & & .252 & .251 & .251 & .251 & .253 & .252 & .253 & .251\\
2 & & & & $U'=2$ & & & & & .252 & .252 & .245 & .253 & .251 & .253 & .251 & .253\\
3 & & & & & & & & & .253 & .241 & .239 & .240 & .253 & .251 & .252 & .252\\
4 & & & & & & & & & .246 & .239 & {\bf .295} & .239 & .245 & .251 & .252 & .251\\
5 & & & & & & & & & .253 & .241 & .239 & .241 & .252 & .251 & .251 & .252\\
6 & & & & & & & & & .252 & .253 & .245 & .252 & .251 & .251 & .252 & .250\\
7 & & & & & & & & & .253 & .252 & .252 & .252 & .251 & .252 & .251 & .252\\
8 & & & & & & & & & & & & & & & &\\
9 & & & & & & & & & & & & & & & &\\
10 &  .253 & .253 & .253 & .253 & .253 & .253 & .253 & .253 & & & & & & & &\\
11 &  .253 & .253 & .253 & .253 & .254 & .253 & .253 & .252 & & & & & & & &\\
12 &  .253 & .254 & .253 & .261 & .253 & .254 & .252 & .253 & & & & & & & &\\
13 &  .253 & .253 & .265 & .248 & .265 & .253 & .254 & .253 & & & & & & & &\\
14 &  .253 & .261 & .248 & {\bf .158} & .248 & .261 & .253 & .253 & & & & & & & &\\
15 &  .254 & .253 & .264 & .248 & .265 & .253 & .253 & .253 & & &$U'=30$& & & & &\\
16 &  .253 & .253 & .253 & .260 & .253 & .253 & .253 & .253 & & & & & & & &\\
\end{tabular}
\end{table}
\begin{multicols}{2}\narrowtext

While these results also follow from the impurity-spin picture,
the small charge gap at the impurity site does have an
impact on the magnon spectrum. 
Within the localized spin picture,
the highest-energy magnon mode corresponding to a local
spin deviation at the impurity site would cost energy 
$4\times J'/2= 16t^2/(U+U')$. However, the highest energy 
in the magnon spectrum is actually seen to be 
$1.05 t$, 
which is substantially smaller than $2J'=1.33 t$. 
This shows the compression effect of the low charge 
gap ($2.75t$) on the magnon spectrum.\cite{spectrum}

In conclusion, using a convenient numerical method for
evaluating transverse spin correlations at the RPA level,
quantum spin-fluctuation corrections to sublattice magnetization 
are obtained for the half-filled Hubbard model in the whole $U/t$ range.
Results in two and three dimensions
are shown to interpolate properly between both the weak
and strong correlation limits, and approach the SWT results 
as $U/t\rightarrow\infty$.
The method is readily extended to other situations of interest
involving defects in the AF, such as vacancies/impurities/disorder. 
Numerical diagonalization for finite lattices,
along with finite-size scaling tested with the pure Hubbard model,
allows for exact treatment of defects at the RPA level.
This is illustrated with a study of the defect-induced 
enhancement/suppression in transverse spin fluctuations 
for spin vacancies and low-$U$ impurities in two dimensions.
While the quantum spin fluctuation correction to sublattice magnetization 
is sharply enhanced by spin vacancies, 
it is strongly suppressed by the low-$U$ impurities, 
although the fluctuation correction is enhanced at the
low$-U$ sites. 

\section*{ACKNOWLEDGMENTS}
Helpful conversations with D. Vollhardt and 
M. Ulmke, and support from the Alexander von Humboldt 
Foundation through a Research Fellowship
at the Universit\"{a}t Augsburg are gratefully acknowledged.

\end{multicols}
\end{document}